\documentclass[proceedings]{JHEP3}
\usepackage{amssymb}
\usepackage{amsfonts}
\usepackage{amsmath}
\usepackage{epsfig,multicol}

\newbox\mybox

\newcommand\fverb{\setbox\mybox=\hbox\bgroup\verb}
\newcommand\fverbdo{\egroup\medskip\noindent\fbox{\unhbox\mybox}\ }
\newcommand\fverbit{\egroup\item[\fbox{\unhbox\mybox}]}
\conference{Non-crystallographic reduction}
\abstract{We apply a recently introduced reduction procedure based on the embedding of 
non-crystallographic Coxeter groups into crystallographic ones to Calogero-Moser
systems. For rational potentials the familiar generalized Calogero Hamiltonian is recovered. 
For the Hamiltonians of trigonometric, hyperbolic and elliptic type, we obtain 
novel integrable dynamical systems with a second potential term which is rescaled by the golden ratio.
We explicitly show for the simplest of these non-crystallographic models how the corresponding 
classical equations of motion can be derived from a Lie algebraic Lax pair based on 
the larger, crystallographic Coxeter group.}
\input{tcilatex}

\title{Non-crystallographic reduction of generalized Calogero-Moser models}
\author{Andreas Fring and Christian Korff \\
Centre for Mathematical Science, City University, \\
Northampton Square, London EC1V 0HB, UK\\
E-mail:  \email{ A.Fring@city.ac.uk, C.Korff@city.ac.uk}}

\begin{document}

\section{Introduction}

The generalized Calogero-Moser models \cite{Cal1,Cal2,Cal3} \cite%
{Suth1,Suth2,Suth3,Suth4} \cite{Moser:75}\ \cite{OP0,OP1,OP2,OP3}\
constitute one of the most prominent and widely studied class of classical
and quantum integrable systems describing $\ell $ particles moving on a
line. In this work we will concentrate on the classical mechanics models.
Given a set of dynamical variables in terms of a coordinate vector $q\in 
\mathbb{R}^{\ell }$ and the canonically conjugate momenta $p\in \mathbb{R}%
^{\ell }$ the classical Calogero-Moser Hamiltonians take the following forms,%
\begin{equation}
H=\frac{p^{2}}{2}+\frac{1}{2}\sum_{\alpha \in \Delta }g_{\alpha
}^{2}V(\alpha \cdot q),\qquad V(u)=\left\{ 
\begin{array}{rr}
1/u^{2}, & \text{rational} \\ 
1/\sin ^{2}u, & \text{trigonometric} \\ 
1/\sinh ^{2}u, & \text{hyperbolic} \\ 
1/\limfunc{sn}^{2}u, & \text{elliptic}%
\end{array}%
\right. \ .  \label{H}
\end{equation}%
The sum in the potential runs over a root system $\Delta \subset \mathbb{R}%
^{\ell }$ associated with a finite Coxeter group $\mathcal{W}$; see for
instance \cite{Hum2} for details. The $g_{\alpha }$ are coupling constants
which must at least coincide on vectors $\alpha $ of the same length, i.e. $%
g_{\alpha }=g_{\beta }$ for $\alpha ^{2}=\beta ^{2},$ if one demands the
Hamiltonian to be invariant under $\mathcal{W}$. In the literature the
elliptic potential in (\ref{H}) is often also expressed in terms of the
Weierstrass $\mathfrak{\wp }$-function, both choices only differ by a
rescaling of the argument and an additive constant \cite{WW} \cite{Cal5}. In
fact, all other types of potentials can be obtained from the elliptic one
through special limits. One often also adds a confining harmonic potential $%
\sim q^{2}$ which we omit from our discussion for simplicity.

If the root system $\Delta $ is crystallographic, i.e. $2(\alpha \cdot \beta
)/\beta ^{2}\in \mathbb{Z}$ for any pair of roots $\alpha ,\beta \in \Delta $%
, the associated Coxeter group $\mathcal{W}$ is connected with semi-simple
Lie algebras \cite{Hum}. Integrability of the Calogero-Moser models (\ref{H}%
) can be proved via the standard technique of Lax pairs $\{L,M\}$ \cite{Lax}%
:\ if the classical equations of motion resulting from (\ref{H}) are
equivalent to the Lax equation $\dot{L}=[L,M],$ the quantities $I_{k}=%
\limfunc{Tr}L^{k}$ are conserved. In contrast to other integrable models
associated with Lie algebras, as for instance (affine) Toda models \cite%
{Wi81} \cite{MOP} \cite{DTfold}, a generic Lie algebraic formulation of the
Lax pair is missing and a variety of alternative approaches have been put
forward in the literature, see e.g. \cite{Cal4} \cite{OP0,OP2} \cite{DP} 
\cite{Sas5,Sas6}. In this article we will only use a Lie algebraic Lax
construction for the $A_{\ell }$ or $su(\ell +1)$ series in order to
exemplify our reduction procedure for the simplest model. However, we stress
that in the aforementioned literature Lax pair constructions have been
carried out for all algebras and for all four types of potentials in (\ref{H}%
).\smallskip

For the non-crystallographic Coxeter groups, $\mathcal{\tilde{W}}%
=I_{2}(m),H_{3},H_{4}$, with root systems $\tilde{\Delta}$ one has in
general $2(\tilde{\alpha}\cdot \tilde{\beta})/\tilde{\beta}^{2}\notin 
\mathbb{Z}$ and the connection with Lie algebras ceases to be valid. Due to
the latter fact the Lax construction now becomes even more difficult and
only the rational potential in (\ref{H}) has been considered using an
alternative formulation based on reflection operators \cite{Sas6}. Other
concepts such as \textquotedblleft exact solvability\textquotedblright\ \cite%
{Tur0,RT}, based on the computation of invariants, also run into problems
for non-rational potentials; see the discussion of an exactly solvable
Sutherland model based on $H_{3}$ in \cite{HR4}.\smallskip

In this letter we overcome these difficulties by introducing for the root
systems $\tilde{\Delta}$ of the non-crystallographic Coxeter groups $%
H_{2}\equiv I_{2}(5),\ H_{3},\ H_{4}$ an extension of the Calogero-Moser
Hamiltonian (\ref{H}) which allows us to tie the proof of integrability for 
\emph{all }four types of potentials to the one of certain crystallographic
groups specified below (\ref{HADE}). Namely, we consider the Hamiltonians%
\begin{equation}
\tilde{H}=\frac{\tilde{p}^{2}}{2}+\frac{\tilde{g}^{2}}{2}\sum_{\tilde{\alpha}%
\in \tilde{\Delta}}\{V(\tilde{\alpha}\cdot \tilde{q})+V(\phi ~\tilde{\alpha}%
\cdot \tilde{q})\},\qquad \phi =\phi ^{2}-1=\frac{1+\sqrt{5}}{2}\ .
\label{newH}
\end{equation}%
The parameter $\phi $ entering the second potential term is the well-known
golden ratio. Clearly, in the case of the rational potential adding the
extra term in (\ref{newH}) amounts to a simple rescaling of the coupling
constant in (\ref{H}) and we have%
\begin{equation}
V(u)=u^{-2}:\qquad \tilde{H}(\tilde{g}^{2})=H((1+\phi ^{-2})\tilde{g}^{2})\ .
\label{Hrat}
\end{equation}%
Thus, we recover the familiar generalization of the Calogero model to
non-crystallographic root systems $\tilde{\Delta}$. For the remaining cases
the insertion of the extra potential term might appear \emph{ad hoc,} at
first sight but we will explain in the text that it occurs naturally in
light of the following embeddings of non-crystallographic Coxeter groups
into crystallographic ones,%
\begin{equation}
H_{2}\hookrightarrow A_{4},\qquad H_{3}\hookrightarrow D_{6},\qquad \text{and%
}\qquad H_{4}\hookrightarrow E_{8}\ .  \label{HADE}
\end{equation}%
Employing a reduction procedure recently introduced in the context of affine
Toda field theory \cite{FKncATFT} the models (\ref{newH}) are obtained from
the Hamiltonians (\ref{H}) of the corresponding crystallographic groups
given in (\ref{HADE}). See also \cite{Pbook} and references therein for
similar reduction procedures. Close analogies also exist with the folding
procedure of crystallographic Coxeter groups linked with simply-laced
algebras into those corresponding to non-simply laced ones (see \cite{DTfold}
\cite{Sasfold} in the context of affine Toda and \cite{BST} for
Calogero-Moser models). However, there are several differences in the
mathematical structure. We will comment on this further in the text.

Our main result in this article is the extension of the non-crystallographic
Calogero-Moser models from the rational case treated so far to
trigonometric, hyperbolic and elliptic potentials. However, our reduction
procedure also puts a new perspective on the familiar rational case, it
enables one to connect the non-crystallographic models to Lie algebras
through the embeddings (\ref{HADE}). The proof of Liouville integrability of
the models (\ref{newH}) can be carried out by employing the structure of the
Lax pairs associated with the crystallographic root systems in (\ref{HADE}).
We explicitly demonstrate this for the simplest model, the one associated
with $H_{2}$, by exploiting a known Lie algebraic Lax pair related to $%
A_{4}\equiv su(5)$, despite the fact that one is dealing with a
non-crystallographic Coxeter group on the level of the Hamiltonian.

The article is structured as follows. In section 2 we review the embeddings (%
\ref{HADE}) and introduce the necessary mathematical formalism for our
reduction procedure. In section 3 it is then explained how to reduce the
crystallographic Hamiltonians (\ref{H}) and the associated equations of
motion to the non-crystallographic systems (\ref{newH}). We address the
question of integrability in section 4 by showing the existence of a Lax
pair for the simplest model associated with $H_{2}$. This Lax pair is
obtained through the reduction of a Lie algebraic pair for $A_{4}$.
Comparison with other Lax pair formulations \cite{Sas5,Sas6} is made in the
appendix. Section 5 contains our conclusions.

\section{Embedding of non-crystallographic into crystallographic Coxeter
groups}

The details of the embeddings (\ref{HADE}) have been presented previously in
the literature \cite{Sher,MP,KRA} and in particular \cite{FKncATFT} which we
follow in our notation. We therefore omit proofs and only present the
necessary formulae for the reduction. Throughout this paper quantities
related to the two types of different Coxeter groups in (\ref{HADE}) will be
distinguished by putting an additional tilde on top of the
non-crystallographic quantities. In light of (\ref{HADE}) we henceforth
limit ourselves to the simply laced case.

Recall \cite{Hum2} that any Coxeter group $\mathcal{W}$ is generated by the
reflections associated with a set of simple roots $\{\alpha _{i}\}\subset
\Delta ,$%
\begin{equation}
\sigma _{i}(x)=x-2\frac{x\cdot \alpha _{i}}{\alpha _{i}\cdot \alpha _{i}}%
\alpha _{i}~\ \qquad \text{for }1\leq i\leq \ell ,~x\in \mathbb{R}^{\ell }.
\label{W}
\end{equation}%
This set of reflections generates the Coxeter group $\mathcal{W}$ subject to
the relations%
\begin{equation}
(\sigma _{i}\sigma _{j})^{m_{ij}}=1,\qquad 1\leq i,j\leq \ell  \label{WW}
\end{equation}%
where the order $m_{ij}\in \mathbb{N}$ of the group elements is defined
through the Cartan matrix $K$,%
\begin{equation}
m_{ij}=\pi /\arccos (-K_{ij}/2),\qquad K_{ij}=\frac{2\alpha _{i}\cdot \alpha
_{j}}{\alpha _{j}\cdot \alpha _{j}}~.  \label{KN}
\end{equation}%
The relations (\ref{W}), (\ref{WW}) and (\ref{KN}) apply also to the
non-crystallographic group $\mathcal{\tilde{W}}$. Since we are only dealing
with root systems $\Delta ,\ \tilde{\Delta}$ where all elements have equal
length, we adopt henceforth the normalization convention $\alpha _{i}^{2}=%
\tilde{\alpha}_{i}^{2}=2$.

Introducing a special labelling of the simple roots $\{\alpha _{i}\}$
depicted in figure 1 allows for combining the different embeddings in (\ref%
{HADE}) into a single formula \cite{FKncATFT},%
\begin{equation}
\mathcal{\tilde{W}}\hookrightarrow \mathcal{W}:\qquad \tilde{\sigma}%
_{i}\mapsto \sigma _{i}\sigma _{i+\tilde{\ell}}=\sigma _{i+\tilde{\ell}%
}\sigma _{i}\qquad \text{for~~}1\leq i\leq \tilde{\ell}~.  \label{emb}
\end{equation}%
Notice that the rank $\tilde{\ell}$ of the non-crystallographic group and
the rank $\ell $ of its crystallographic counterpart in (\ref{HADE}) are
always related by a factor two, $\ell =2\tilde{\ell}$. Furthermore, our
labelling of the simple roots is such that the roots $\alpha _{i}$, $\alpha
_{i+\tilde{\ell}}$ are always orthogonal whence the associated reflections
commute. The embedding (\ref{emb}) is to be understood in the sense of a
group homomorphism, i.e. it preserves the Coxeter relations (\ref{WW}%
).\smallskip

\unitlength=0.6500000pt 
\begin{picture}(300.0,70.00)(150.00,125.00)

\put(140.00,150.00){\makebox(0.00,0.00){$A_{4}:$}}
\put(389.00,165.00){\makebox(0.00,0.00){${\alpha}_{2}$}}
\put(348.0,165.00){\makebox(0.00,0.00){${\alpha}_{3}$}}
\put(296.00,165.00){\makebox(0.00,0.00){${\alpha}_4$}}
\put(255.00,165.00){\makebox(0.00,0.00){${\alpha}_1$}}

\put(350.00,150.00){\line(1,0){30.00}}
\put(330.00,150.00){\line(1,0){10.00}}
\put(300.00,150.00){\line(1,0){30.00}}
\put(260.00,150.00){\line(1,0){30.00}}

\put(385.00,150.00){\circle*{10.00}}
\put(345.00,150.00){\circle*{10.00}}
\put(255.00,150.00){\circle*{10.00}}
\put(295.00,150.00){\circle*{10.00}}
\put(450.00,155.00){$\omega$}
\put(440.00,147.00){$ \longrightarrow $}

\put(638.00,165.00){\makebox(0.00,0.00){${\tilde{\alpha}}_2$}}
\put(597.00,165.00){\makebox(0.00,0.00){${\phi \tilde{\alpha}}_1$}}

\put(546.00,165.00){\makebox(0.00,0.00){${\phi \tilde{\alpha}}_2$}}
\put(505.00,165.00){\makebox(0.00,0.00){${ \tilde{ \alpha}}_1$}}

\put(600.00,150.00){\line(1,0){30.00}}
\put(580.00,150.00){\line(1,0){10.00}}
\put(550.00,150.00){\line(1,0){30.00}}
\put(510.00,150.00){\line(1,0){30.00}}

\put(635.00,150.00){\circle*{10.00}}
\put(595.00,150.00){\circle*{10.00}}
\put(505.00,150.00){\circle*{10.00}}
\put(545.00,150.00){\circle*{10.00}}
\end{picture}

\unitlength=0.6500000pt 
\begin{picture}(237.92,150.00)(220.00,65.00)

\put(210.00,150.00){\makebox(0.00,0.00){$D_{6}:$}}
\put(432.0,107.17){\makebox(0.00,0.00){$\alpha_{3}$}}
\put(437.92,184.17){\makebox(0.00,0.00){${\alpha}_{4}$}}
\put(408.17,149.59){\makebox(0.00,0.00){${\alpha}_{5}$}}
\put(347.50,165.00){\makebox(0.00,0.00){${\alpha}_{6}$}}
\put(296.00,165.00){\makebox(0.00,0.00){${\alpha}_2$}}
\put(255.00,165.00){\makebox(0.00,0.00){${\alpha}_1$}}
\put(389.00,146.33){\line(1,-1){22.67}}
\put(411.67,176.67){\line(-1,-1){23.00}}
\put(350.00,150.00){\line(1,0){30.00}}
\put(330.00,150.00){\line(1,0){10.00}}
\put(300.00,150.00){\line(1,0){30.00}}
\put(260.00,150.00){\line(1,0){30.00}}
\put(415.00,120.00){\circle*{10.00}}
\put(415.00,180.00){\circle*{10.00}}
\put(385.00,150.00){\circle*{10.00}}
\put(345.00,150.00){\circle*{10.00}}
\put(255.00,150.00){\circle*{10.00}}
\put(295.00,150.00){\circle*{10.00}}
\put(500.00,155.00){$\omega$}
\put(490.00,147.00){$ \longrightarrow $}
\put(732.0,107.17){\makebox(0.00,0.00){ $ {\tilde{\alpha}}_{3}$ }}
\put(737.92,184.17){\makebox(0.00,0.00){$ \phi {\tilde{\alpha}}_{1}$}}
\put(708.17,149.59){\makebox(0.00,0.00){ $ \phi {\tilde{\alpha}}_{2}$}}
\put(647.50,165.00){\makebox(0.00,0.00){ $ \phi {\tilde{\alpha}}_{3}$ }}
\put(596.00,165.00){\makebox(0.00,0.00){${\tilde{\alpha}}_2$}}
\put(555.00,165.00){\makebox(0.00,0.00){${ \tilde{ \alpha}}_1$}}
\put(689.00,146.33){\line(1,-1){22.67}}
\put(711.67,176.67){\line(-1,-1){23.00}}
\put(650.00,150.00){\line(1,0){30.00}}
\put(630.00,150.00){\line(1,0){10.00}}
\put(600.00,150.00){\line(1,0){30.00}}
\put(560.00,150.00){\line(1,0){30.00}}
\put(715.00,120.00){\circle*{10.00}}
\put(715.00,180.00){\circle*{10.00}}
\put(685.00,150.00){\circle*{10.00}}
\put(645.00,150.00){\circle*{10.00}}
\put(555.00,150.00){\circle*{10.00}}
\put(595.00,150.00){\circle*{10.00}}
\end{picture}

\unitlength=0.6500000pt 
\begin{picture}(370.00,107.58)(180.00,0.00)
\put(170.00,38.00){\makebox(0.00,0.00){$E_8:$}}
\put(445.00,52.00){\makebox(0.00,0.00){$\alpha_1$}}
\put(405.00,52.00){\makebox(0.00,0.00){$\alpha_2$}}
\put(365.00,52.00){\makebox(0.00,0.00){$\alpha_3$}}
\put(325.00,52.00){\makebox(0.00,0.00){$\alpha_8$}}
\put(296.25,52.17){\makebox(0.00,0.00){${\alpha}_7$}}
\put(285.00,93.00){\makebox(0.00,0.00){${\alpha}_4$}}
\put(245.00,52.00){\makebox(0.00,0.00){${\alpha}_6$}}
\put(206.00,52.00){\makebox(0.00,0.00){${\alpha}_5$}}
\put(410.00,38.00){\line(1,0){30.00}}
\put(370.00,38.00){\line(1,0){30.00}}
\put(330.00,38.00){\line(1,0){30.00}}
\put(285.33,72.67){\line(0,-1){31.00}}
\put(290.00,38.00){\line(1,0){30.00}}
\put(250.00,38.00){\line(1,0){30.00}}
\put(210.00,38.00){\line(1,0){30.00}}
\put(285.00,78.00){\circle*{10.00}}
\put(245.00,38.00){\circle*{10.00}}
\put(325.00,38.00){\circle*{10.00}}
\put(365.00,38.00){\circle*{10.00}}
\put(405.00,38.00){\circle*{10.00}}
\put(445.00,38.00){\circle*{10.00}}
\put(285.00,38.00){\circle*{10.00}}
\put(205.00,38.00){\circle*{10.00}}
\put(500.00,45.00){$\omega$}
\put(490.00,35.00){$ \longrightarrow $}
\put(795.00,52.00){\makebox(0.00,0.00){ $\tilde{\alpha}_1$ }}
\put(755.00,52.00){\makebox(0.00,0.00){ $\tilde{\alpha}_2$ }}
\put(715.00,52.00){\makebox(0.00,0.00){ $\tilde{\alpha}_3$ }}
\put(685.00,52.00){\makebox(0.00,0.00){ $ \phi \tilde{\alpha}_4$ }}
\put(651.25,52.17){\makebox(0.00,0.00){ $\phi \tilde{\alpha}_3$ }}
\put(635.00,93.00){\makebox(0.00,0.00){$\tilde{\alpha}_4$ }}
\put(595.00,52.00){\makebox(0.00,0.00){$ \phi \tilde{\alpha}_2$ }}
\put(556.00,52.00){\makebox(0.00,0.00){ $\phi \tilde{\alpha}_1$ }}
\put(760.00,38.00){\line(1,0){30.00}}
\put(720.00,38.00){\line(1,0){30.00}}
\put(680.00,38.00){\line(1,0){30.00}}
\put(635.33,72.67){\line(0,-1){31.00}}
\put(640.00,38.00){\line(1,0){30.00}}
\put(600.00,38.00){\line(1,0){30.00}}
\put(560.00,38.00){\line(1,0){30.00}}
\put(635.00,78.00){\circle*{10.00}}
\put(595.00,38.00){\circle*{10.00}}
\put(675.00,38.00){\circle*{10.00}}
\put(715.00,38.00){\circle*{10.00}}
\put(755.00,38.00){\circle*{10.00}}
\put(795.00,38.00){\circle*{10.00}}
\put(635.00,38.00){\circle*{10.00}}
\put(555.00,38.00){\circle*{10.00}}
\end{picture}\smallskip

\begin{center}
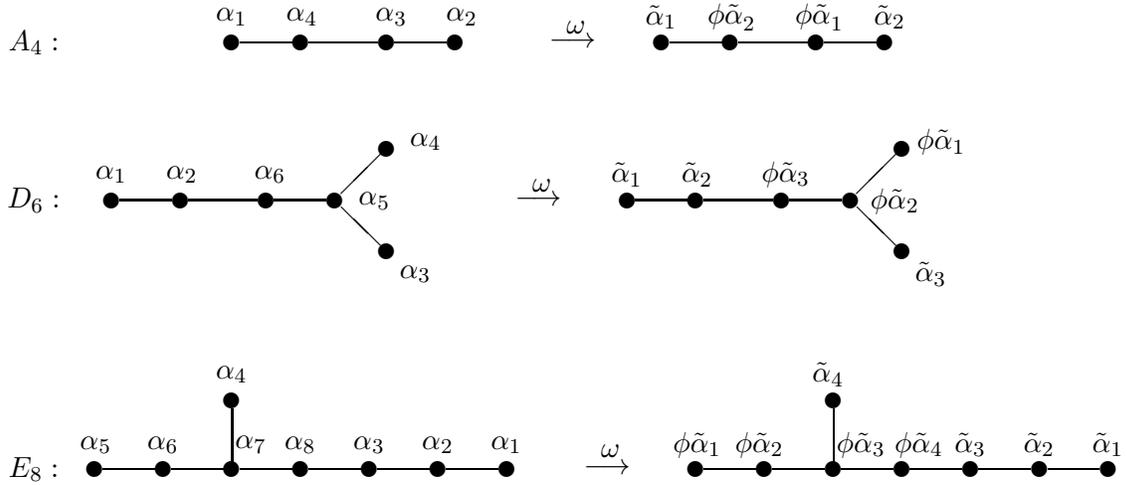
{\small Figure 1: Coxeter graphs, root labelling and the map (\ref{om1}) for
the Coxeter groups in (\ref{HADE}). }
\end{center}

In order to realize (\ref{emb}) in the context of the Calogero-Moser models (%
\ref{H}), we need to know how this embedding manifests itself on the level
of the corresponding root spaces $\Delta $ and $\tilde{\Delta}$ which are
left invariant by $\mathcal{W}$ and$\ \mathcal{\tilde{W}}$, respectively.
This is achieved by defining a pair of maps%
\begin{equation}
\omega :~\Delta \rightarrow \tilde{\Delta}\cup \phi \tilde{\Delta}\qquad 
\text{and}\qquad \tilde{\omega}:\tilde{\Delta}\rightarrow \Delta \oplus \phi
\Delta ,  \label{defom}
\end{equation}%
which intertwine the embedding (\ref{emb}), i.e. 
\begin{equation}
\tilde{\sigma}_{i}\omega =\omega \sigma _{i}\sigma _{i+\tilde{\ell}}\qquad 
\text{and}\qquad \tilde{\omega}\tilde{\sigma}_{i}=\sigma _{i}\sigma _{i+%
\tilde{\ell}}\tilde{\omega}\;.  \label{ss}
\end{equation}%
The first map $\omega $ has been previously considered in the literature,
see e.g. \cite{MP} \cite{FKncATFT}, and is defined as follows%
\begin{equation}
\alpha _{i}\mapsto \omega (\alpha _{i})=\left\{ 
\begin{array}{ll}
\tilde{\alpha}_{i} & \text{for\quad }1\leq i\leq \tilde{\ell}=\ell /2 \\ 
\phi \tilde{\alpha}_{i-\tilde{\ell}}\qquad & \text{for\quad }\tilde{\ell}%
<i\leq \ell .%
\end{array}%
\right.  \label{om1}
\end{equation}%
The second map $\tilde{\omega},$ introduced in \cite{FKncATFT} and paramount
to our reduction procedure, realizes the simple root system $\{\tilde{\alpha}%
_{i}\}$ of the non-crystallographic Coxeter group $\mathcal{\tilde{W}}$ in $%
\mathbb{R}^{\ell }$ by identifying%
\begin{equation}
\tilde{\alpha}_{i}\mapsto \tilde{\omega}(\tilde{\alpha}_{i})=\alpha
_{i}+\phi \alpha _{i+\tilde{\ell}}\qquad \text{for~~}1\leq i\leq \tilde{\ell}%
~.  \label{om2}
\end{equation}%
The images of the non-crystallographic roots have now length $\tilde{\omega}(%
\tilde{\alpha}_{i})^{2}=2(1+\phi ^{2})$ according to our earlier convention $%
\tilde{\alpha}_{i}^{2}=2$. Thus $\tilde{\omega}$ preserves the inner product
only up to a factor $(1+\phi ^{2})$.

As the simple roots $\{\alpha _{i}\}$ and $\{\tilde{\alpha}_{i}\}$ are
linearly independent the maps (\ref{om1}), (\ref{om2}) can be linearly
extended to the whole vector spaces $\mathbb{R}^{\ell }$ respectively $%
\mathbb{R}^{\tilde{\ell}}$. We will make use of this fact when reducing the
crystallographic Calogero-Moser systems to non-crystallographic ones below.
Note that the defining relations (\ref{om1}), (\ref{om2}) also apply to the
fundamental weights \cite{FKncATFT}.

Using the pair $\omega ,\tilde{\omega}$ we are in the position to relate
inner products in $\tilde{\Delta}$ to inner products in $\Delta $ by means
of the identity%
\begin{equation}
\omega (\alpha _{i})\cdot \tilde{\alpha}_{j}=\alpha _{i}\cdot \tilde{\omega}(%
\tilde{\alpha}_{j})\qquad \text{for~~}1\leq j\leq \tilde{\ell}~,1\leq i\leq
\ell ~.  \label{oot}
\end{equation}%
From this relationship as well as (\ref{defom}), (\ref{ss}) we infer that $%
\tilde{\omega}$ plays the role of a \textquotedblleft
quasi-inverse\textquotedblright\ to the map $\omega $, in fact we have that%
\begin{equation}
\omega \tilde{\omega}=(1+\phi ^{2})~\mathbb{I\qquad }\text{and\qquad }\tilde{%
\omega}\omega =\left( 
\begin{array}{cc}
\mathbb{I} & \phi \mathbb{I} \\ 
\phi \mathbb{I} & \phi ^{2}\mathbb{I}%
\end{array}%
\right) ,  \label{ooi}
\end{equation}%
with $\mathbb{I}$ denoting the $\tilde{\ell}\times \tilde{\ell}$ identity
matrix. As an immediate consequence of (\ref{oot}) we obtain a crucial
relationship between the non-crystallographic Cartan matrix $\tilde{K}$ and
the crystallographic one $K$. Namely, introducing the $\tilde{\ell}\times 
\tilde{\ell}$ matrices $\kappa $ and $\hat{\kappa}$ through the following
block decomposition of the crystallographic Cartan matrix 
\begin{equation}
K=\left( 
\begin{array}{cc}
\kappa & \hat{\kappa} \\ 
\hat{\kappa} & \kappa +\hat{\kappa}%
\end{array}%
\right) ,  \label{kK}
\end{equation}%
we have the matrix equation%
\begin{equation}
\tilde{K}=\kappa +\phi \hat{\kappa}=\phi ^{-1}\hat{\kappa}+\kappa +\hat{%
\kappa}\ .  \label{KK}
\end{equation}%
Employing the definitions (\ref{om1}), (\ref{om2}) together with the
identities (\ref{oot}), (\ref{KK}) the intertwining relations (\ref{ss}) now
follow from a straightforward computation. Similar identities also hold for
the inverse Cartan matrices \cite{FKncATFT}.

\section{Reduction of crystallographic Calogero-Moser models}

Having introduced the necessary mathematical set-up we are now in the
position to introduce our reduction map. We start from a dynamical system
defined in terms of the Hamiltonian (\ref{H}) based on any of the three
crystallographic Coxeter groups in (\ref{HADE}). Such a system depends on $%
\ell $-independent dynamical variables $q=(q_{1},...,q_{\ell })$ and $\ell $%
-independent conjugate momenta $p=(p_{1},...,p_{\ell })$. We now replace
this set of variables by a new one which only contains $\tilde{\ell}$%
-independent coordinates and $\tilde{\ell}$-independent momenta by defining
the following reduction map $\mu ,$%
\begin{equation}
(q,p)\rightarrow (\mu (q),\mu (p)):=(\tilde{\omega}(\tilde{q}),\tilde{\omega}%
(\tilde{p}))\ .  \label{mu}
\end{equation}%
Here the action of $\tilde{\omega}$ on the simple roots $\tilde{\alpha}_{i}$
is defined in (\ref{om2}). The vectors $\tilde{q}=(\tilde{q}_{1},...,\tilde{q%
}_{\tilde{\ell}})$ and$\;\tilde{p}=(\tilde{p}_{1},...,\tilde{p}_{\tilde{\ell}%
})$ in the Euclidean basis will become the dynamical variables with respect
to the non-crystallographic Hamiltonian (\ref{newH}) resulting from the
reduced Hamiltonian%
\begin{equation}
H(q,p)\overset{\mu }{\rightarrow }H^{\text{red}}:=H(\tilde{\omega}(\tilde{q}%
),\tilde{\omega}(\tilde{p}))\ .  \label{redH}
\end{equation}%
Our particular choice for the definition of the reduction map (\ref{mu}) in
terms of the map $\tilde{\omega}$ will allow us to discuss the reduction
procedure without making reference to a specific representation of the root
spaces $\Delta $ and $\tilde{\Delta}$. By exploting the identity (\ref{oot})
the reduction (\ref{mu}) can be carried out in terms of the root systems
instead of the dynamical variables and momenta.

Let us further motivate (\ref{mu}) by comparing it to the reduction when
folding a simply laced Lie algebra by a non-trivial Dynkin diagram
automorphism $\tau $ to a non-simply laced algebra \cite{DTfold}. In that
context the reduction occurs when the coordinates $q$ and momenta$\ p$ are
projected onto the invariant subspaces under $\tau $. This decreases the
number of independent variables. The reduced or folded Calogero-Moser
Hamiltonian \cite{BST} is then obtained by inserting the projected variables
into the original \textquotedblleft simply laced\textquotedblright\
Hamiltonian (\ref{H}) and rewriting it in terms of the $\tau $-invariant
root subspace $\Delta _{\tau }\subset \Delta $. The latter can be identified
with the root system $\Delta ^{\text{ns}}$ of a the non-simply laced
algebra. We will comment further on this analogy below, see (\ref{nsH}).

We proceed here analogously and now explain how the reduced Hamiltonian (\ref%
{redH}) can be expressed in terms of the non-crystallographic root system $%
\tilde{\Delta}$ only.

\subsection{The reduced Hamiltonian}

As may be seen in (\ref{HADE}) all the Lie algebras relevant to our
reduction procedure are simply laced and we set $g_{\alpha }=g$ in (\ref{H}%
). Then the reduced Hamiltonian (\ref{redH}) can be rewritten as follows,%
\begin{eqnarray}
2H^{\text{red}} &=&\tilde{\omega}(\tilde{p})^{2}+g^{2}\tsum_{\alpha \in
\Delta }V(\alpha \cdot \tilde{\omega}(\tilde{q}))  \label{newH0} \\
&=&(1+\phi ^{2})\tilde{p}^{2}+g^{2}\tsum_{\alpha \in \Delta }V(\omega
(\alpha )\cdot \tilde{q})  \notag \\
&=&(1+\phi ^{2})\left\{ \tilde{p}^{2}+\tilde{g}^{2}\tsum_{\tilde{\alpha}\in 
\tilde{\Delta}}V(\tilde{\alpha}\cdot \tilde{q})+\tilde{g}^{2}\tsum_{\tilde{%
\alpha}\in \tilde{\Delta}}V(\phi ~\tilde{\alpha}\cdot \tilde{q})\right\}
=2(1+\phi ^{2})\tilde{H}\ .  \label{newH2}
\end{eqnarray}%
In the last line all data belong to the non-crystallographic root system and
we have arrived at (\ref{newH}). Let us first explain the reduction of the
potential term. To obtain the second line we have used the inner product
identity (\ref{oot}) which replaces crystallographic roots by
non-crystallographic ones. Exploiting that the map $\omega $ defined in (\ref%
{om1}) is surjective we arrive at the third line (\ref{newH2}). Here the sum
over the crystallographic root system $\Delta $ is now replaced by sums over
the two copies of the non-crystallographic root space $\tilde{\Delta}$
appearing in the target space of $\omega $; compare with (\ref{defom}). In
the last line we have also defined a rescaled coupling constant by setting 
\begin{equation}
g=\sqrt{1+\phi ^{2}}~\tilde{g}~.  \label{ggt}
\end{equation}%
This scaling factor arises from the kinetic energy term and is due to the
fact that $\tilde{\omega}$ is not an isometry. Expanding $\tilde{p}=\sum_{i}%
\tilde{r}^{i}\tilde{\alpha}_{i}$ we compute%
\begin{equation}
\tilde{\omega}(\tilde{p})\cdot \tilde{\omega}(\tilde{p})=(1+\phi
^{2})\tsum_{i,j=1}^{\tilde{\ell}}\tilde{r}^{i}(\kappa _{ij}+\phi \hat{\kappa}%
_{ij})\tilde{r}^{j}=(1+\phi ^{2})\tsum_{i,j=1}^{\tilde{\ell}}\tilde{r}^{i}%
\tilde{K}_{ij}\tilde{r}^{j}=(1+\phi ^{2})\tilde{p}^{2},
\end{equation}%
where $\kappa ,\hat{\kappa}$ have been defined in (\ref{kK}). Furthermore,
we have used the normalization convention $\tilde{\alpha}_{i}^{2}=\alpha
_{i}^{2}=2$.

\subsection{Invariance under the non-crystallographic Coxeter group}

It is apparent from the explicit form of the Hamiltonian (\ref{newH2}) that
the reduced Calogero-Moser model is invariant under the non-crystallographic
Coxeter group $\mathcal{\tilde{W}}$. Employing the intertwining property (%
\ref{ss}) and the fact that Coxeter transformations preserve the inner
product, we see that the action of $\mathcal{\tilde{W}}$ in the
\textquotedblleft crystallographic variant\textquotedblright\ (\ref{newH0})
of the new Hamiltonian (\ref{newH}) is realized through the embedding (\ref%
{emb}). For instance, we have for the potential%
\begin{equation}
\sum_{\alpha \in \Delta }V(\tilde{\omega}(\tilde{\sigma}_{i}\tilde{q})\cdot
\alpha )=\sum_{\alpha \in \Delta }V(\sigma _{i}\sigma _{i+\tilde{\ell}}%
\tilde{\omega}(\tilde{q})\cdot \alpha )=\sum_{\alpha \in \Delta }V(\tilde{%
\omega}(\tilde{q})\cdot \alpha )\ .
\end{equation}%
A similar identity holds for the kinetic term. We can use this fact to show
that the coupling constants in front of the two potential terms in (\ref%
{newH2}) can be chosen independently without violating invariance under the
non-crystallographic Coxeter group $\mathcal{\tilde{W}}$. This is apparent
from the variant (\ref{newH2}), but as a preparatory step for the reduction
of a crystallographic Lax pair below it is instructive to directly verify
this also in terms of the reduced Hamiltonian (\ref{newH0}).

First we need to split the crystallographic root system $\Delta $ into the
following disjoint subsets,%
\begin{equation}
\Delta =\Delta ^{\prime }\cup \Delta ^{\prime \prime }\ \quad \text{with\ }%
\quad \omega (\Delta ^{\prime })=\tilde{\Delta}\quad \text{and}\quad \omega
(\Delta ^{\prime \prime })=\phi \tilde{\Delta}\ .  \label{DDD}
\end{equation}%
In order to see that these sets are indeed disjoint, note that for any root $%
\tilde{\alpha}\in \tilde{\Delta}$ the vector $\phi \tilde{\alpha}\notin 
\tilde{\Delta}$. Otherwise there had to be a group element $\tilde{w}\in 
\mathcal{\tilde{W}}$ which maps $\tilde{\alpha}$ into $\phi \tilde{\alpha}$,
as the action of the Coxeter group exhausts the entire root space. If such
an element would exist, we had the identity $\tilde{w}(\tilde{\alpha}%
)^{2}=\phi ^{2}\tilde{\alpha}^{2}\neq 2$ which contradicts the fact that the
Coxeter group $\mathcal{\tilde{W}}$ preserves the inner product. We can
therefore conclude that $\tilde{\Delta}\cap \phi \tilde{\Delta}=\varnothing $
and therefore $\Delta ^{\prime }\cap \Delta ^{\prime \prime }=\varnothing $.
This then implies that $\Delta ^{\prime }$ and $\Delta ^{\prime \prime }$
must be left invariant under the action of the non-crystallographic Coxeter
group with respect to the embedding (\ref{emb}). Namely, according to (\ref%
{ss}) we have for any root $\alpha ^{\prime }\in \Delta ^{\prime }$ that 
\begin{equation*}
\tilde{\sigma}_{i}\omega (\alpha ^{\prime })=\omega (\sigma _{i}\sigma _{i+%
\tilde{\ell}}\alpha ^{\prime })\in \tilde{\Delta}
\end{equation*}%
which entails that $\sigma _{i}\sigma _{i+\tilde{\ell}}\Delta ^{\prime
}=\Delta ^{\prime }$ for all $i=1,...,\tilde{\ell}$. A similar argument
holds for $\Delta ^{\prime \prime }$. Taking invariance under the Coxeter
group $\mathcal{\tilde{W}}$ as a guiding principle, we can therefore
generalize (\ref{newH0}) by introducing the following modified reduced
crystallographic Hamiltonian,%
\begin{equation}
H^{\text{red}}=\frac{\tilde{\omega}(\tilde{p})^{2}}{2}+\frac{g_{1}^{2}}{2}%
\sum_{\alpha ^{\prime }\in \Delta ^{\prime }}V(\alpha ^{\prime }\cdot \tilde{%
\omega}(\tilde{q}))+\frac{g_{2}^{2}}{2}\sum_{\alpha ^{\prime \prime }\in
\Delta ^{\prime \prime }}V(\alpha ^{\prime \prime }\cdot \tilde{\omega}(%
\tilde{q})),  \label{newH02}
\end{equation}%
where $g_{1},g_{2}$ are now arbitrary. The appearance of an additional free
coupling constant in the reduction is very reminiscent of the folding
procedure \cite{DTfold} in the context of Calogero-Moser models \cite{BST}
already mentioned previously.

\subsection{Comparison with folding}

The structure of the Calogero-Moser Hamiltonian (\ref{newH02}) is similar to
the one obtained by folding a simply laced root system $\Delta $ into a
non-simply laced one $\Delta ^{\text{ns}}$ via a Dynkin diagram automorphism 
$\tau $. The potential term in the \textquotedblleft
folded\textquotedblright\ Hamiltonian \cite{BST} also splits into two parts,%
\begin{equation}
H^{\text{ns}}=\frac{p^{2}}{2}+\frac{g_{s}^{2}}{2}\sum_{\alpha \in \Delta
_{s}}V(\alpha \cdot q)+\frac{g_{l}^{2}}{2}\sum_{\alpha \in \Delta
_{l}}V(\alpha \cdot q),  \label{nsH}
\end{equation}%
one running over the short roots $\Delta _{s},$\ the other over the long
roots $\Delta _{l}$, each constituting an independent Weyl group orbit. Note
the absence of the scaling factor in the second term in comparison to (\ref%
{newH02}).\footnote{%
We only mention here the case which has been referred to as
\textquotedblleft untwisted\textquotedblright\ in \cite{BST}, i.e. $\tau $
is an automorphism related to the non extended Dynkin diagram; see \cite{BST}
\cite{Sas8} for other possibilities.} Similar as in our initial calculation
leading to (\ref{newH2}) the coupling constants $g_{s},\ g_{l}$ are not
independent but related by $|\tau |$ due to the folding procedure. However,
outside the framework of folding one can often choose the two couplings
independently without violating integrability \cite{OP0,OP2,Pbook} \cite%
{Sas5,BST}. An exception in the framework of Lie algebraic Lax pairs is the $%
B_{\ell }$ series \cite{Pbook} and $G_{2}$ \cite{AFNM}. We will verify below
whether we can retain in the present context the two independent couplings
in (\ref{newH02}) for the reduction of a Lie algebraic Lax pair.

\subsection{The equations of motion}

Before the construction of a Lax pair we first apply our reduction procedure
to the classical equations of motion. Let $\nabla _{q}$ denote the gradient
operator with respect to the Euclidean basis in $q$-space~($\cong \mathbb{R}%
^{\ell }$). Then the equations of motion originating from the
crystallographic variant (\ref{newH02}) of the Hamiltonian are%
\begin{equation}
\tilde{\omega}(\tilde{p})=\tilde{\omega}(\overset{~.}{\tilde{q}})\qquad 
\text{and\qquad }\tilde{\omega}(\overset{~.}{\tilde{p}})=-\nabla _{q}H^{%
\text{red}}=-\frac{1}{2}\sum_{\alpha \in \Delta }\alpha ~g_{\alpha
}^{2}V^{\prime }(\alpha \cdot \tilde{\omega}(\tilde{q}))\ ,  \label{eqm0}
\end{equation}%
where we set $g_{\alpha }=g_{1}$ for $\alpha \in \Delta ^{\prime }$ and $%
g_{\alpha }=g_{2}$ for $\alpha \in \Delta ^{\prime \prime }$. Acting on both
sides of these two equations with the map $\omega $ defined in (\ref{om1})
together with the identities (\ref{oot}), (\ref{ooi}) we obtain the reduced
system%
\begin{equation}
\tilde{p}=\overset{~.}{\tilde{q}}\qquad \text{and\qquad }\overset{~.}{\tilde{%
p}}=-\nabla _{\tilde{q}}\tilde{H}=-\frac{1}{2}\sum_{\tilde{\alpha}\in \tilde{%
\Delta}}\left\{ \tilde{\alpha}~\tilde{g}_{1}^{2}V^{\prime }(\tilde{\alpha}%
\cdot \tilde{q})+\phi \tilde{\alpha}~\tilde{g}_{2}^{2}V^{\prime }(\phi 
\tilde{\alpha}\cdot \tilde{q})\right\}  \label{eqm}
\end{equation}%
corresponding to the non-crystallographic Hamiltonian (\ref{newH2}). Here $%
\tilde{g}_{i}=g_{i}/(1+\phi ^{2})^{1/2},\ i=1,2$ and $\nabla _{\tilde{q}}$
is now the gradient operator with respect to the Euclidean basis in $\tilde{q%
}$-space~($\cong \mathbb{R}^{\tilde{\ell}}$). Notice that the system (\ref%
{eqm0}) is more restrictive than (\ref{eqm}), i.e. any solution to (\ref%
{eqm0}) yields a solution of (\ref{eqm}) but the converse is not necessarily
true. To see this, one can apply the map $\tilde{\omega}$ on both sides of (%
\ref{eqm}) using that $\tilde{\omega}(\tilde{\alpha})=\alpha ^{\prime }+\phi
\alpha ^{\prime \prime }$ with $\tilde{\alpha}=\omega (\alpha ^{\prime
})=\phi ^{-1}\omega (\alpha ^{\prime \prime })$. A similar observation
applies in the context of folding.

The first crucial step to show integrability of our reduced systems is to
show that (\ref{eqm}) can be equivalently formulated in terms of a Lax pair.
In this context our ability to express the non-crystallographic Hamiltonian (%
\ref{newH2}) and the equations of motion (\ref{eqm}) in crystallographic
terms, (\ref{newH0}) and (\ref{eqm0}), will be essential.

\section{A Lie algebraic Lax pair for the $H_{2}$ model}

As pointed out in the introduction there is no generic Lie algebraic
formulation for the Lax pairs of the generalized Calogero-Moser models.
Instead a variety of different constructions for Lax pairs has been put
forward in the literature, see e.g. \cite{Cal4} \cite{OP0,OP2} \cite{Pbook} 
\cite{DP} \cite{Sas5,Sas6}, which will not be discussed in detail. We focus
on the simplest model associated with $H_{2}$, where according to (\ref{HADE}%
) the corresponding crystallographic system is related to $A_{4}$
respectively $su(5)$. The original construction of the Lax pair for the $%
A_{\ell }$ series goes back to Calogero \cite{Cal4}. We shall adopt here its
formulation in the Cartan-Weyl basis (as it can be found for instance in 
\cite{OP0,Pbook} \cite{Braden:1997}), since in this setting the computation
is more general. For the moment we keep the rank $\ell $ arbitrary, such
that one can easily adopt our discussion to the cases in (\ref{HADE})
omitted here. We shall specialize to the relevant case $\ell =4$ below.

Consider the Cartan-Weyl basis defined through the commutation relations
(e.g. \cite{Hum})%
\begin{equation}
\lbrack H_{i},E_{\alpha }]=\alpha ^{i}E_{\alpha },\qquad \lbrack E_{\alpha
},E_{-\alpha }]=\alpha \cdot H,\qquad \lbrack E_{\alpha },E_{\beta
}]=\varepsilon _{\alpha ,\beta }~E_{\alpha +\beta }\ .  \label{CartanWeyl}
\end{equation}%
In the last commutator it is understood that $\alpha \neq -\beta $ and $%
\varepsilon _{\alpha ,\beta }=0$ whenever $\alpha +\beta $ is not an element
of the root space $\Delta $. The compatible choice of the trace is%
\begin{equation}
\limfunc{Tr}(H_{i}H_{j})=\delta _{ij}\qquad \text{and}\qquad \limfunc{Tr}%
(E_{\alpha }E_{-\alpha })=1,  \label{norm}
\end{equation}%
which implies together with our previous convention $\alpha ^{2}=2$ that the
structure constants $\varepsilon _{\alpha ,\beta }$ only assume the values $%
0 $, $\pm 1$. In addition, they satisfy the following general identities%
\begin{equation}
\varepsilon _{\alpha ,\beta }=-\varepsilon _{\beta ,\alpha }=-\varepsilon
_{-\alpha -\beta ,\beta }\ .  \label{struc}
\end{equation}%
We require $E_{\alpha }^{\dag }=E_{-\alpha }$ which implies the further
constraint $\varepsilon _{\alpha ,\beta }=-\varepsilon _{-\alpha ,-\beta }$.
The Lax pair is now expressed in terms of the Cartan-Weyl basis as follows,%
\begin{equation}
L=p\cdot H+i\sum_{\alpha \in \Delta }g_{\alpha }x(\alpha \cdot q)E_{\alpha
}\qquad \text{and}\qquad M=z\cdot H+i\sum_{\alpha \in \Delta }g_{\alpha
}x^{\prime }(\alpha \cdot q)E_{\alpha }\ .  \label{LM}
\end{equation}%
Here we have once more introduced the root dependent coupling constants $%
g_{\alpha }=g_{1}$ for $\alpha \in \Delta ^{\prime }$ and $g_{\alpha }=g_{2}$
for $\alpha \in \Delta ^{\prime \prime }$ in light of our reduced
Hamiltonian (\ref{newH02}). For the original $A_{\ell }$ Calogero-Moser
model one has $g_{\alpha }=g$. The vector $z=z(q)\in \mathbb{R}^{\ell }$ in (%
\ref{LM}) will be specified momentarily. Let us first define the coefficient
function $x=x(u)$, which can take one of the following forms for the various
types of potentials in (\ref{H}) \cite{Cal5} \cite{OP0,OP2,Pbook},%
\begin{equation}
x(u)=\left\{ 
\begin{array}{cc}
1/u, & \text{rational} \\ 
1/\sin u, & \text{trigonometric} \\ 
1/\sinh u, & \text{hyperbolic} \\ 
1/\limfunc{sn}u, & \text{elliptic}%
\end{array}%
\right. \ .  \label{x}
\end{equation}%
There are other possible choices for the coefficient functions \cite{Pbook}
which can also depend on a spectral parameter \cite{Kr80} \cite{DP}\ \cite%
{Sas6}. For our disucssion of the reduction procedure we picked the present
ones, because they are the simplest, but our results do also apply to the
more general cases. The coefficient functions (\ref{x}) satisfy a number of
identities%
\begin{equation}
x(u)=-x(-u),\qquad x(u)x(-u)=-V(u),  \label{xxV}
\end{equation}%
and crucially \cite{Cal5} \cite{OP0}%
\begin{equation}
x(u)x^{\prime }(w)-x^{\prime }(u)x(w)=\left[ V(u)-V(w)\right] ~x(u+w)\ .
\label{xxV2}
\end{equation}%
Using the first two relations (\ref{xxV}) one shows that%
\begin{equation}
\limfunc{Tr}L^{2}=p^{2}-\sum_{\alpha \in \Delta }x_{\alpha }x_{-\alpha }=2H
\label{L2H}
\end{equation}%
with $x_{\alpha }$ being shorthand notation for $x_{\alpha }(q)=g_{\alpha
}x(\alpha \cdot q)$. The third identity (\ref{xxV2}) comes into play when
showing the equivalence of the classical equations of motion to the
aforementioned Lax equation%
\begin{equation}
\dot{L}=[L,M]\ .  \label{Lax}
\end{equation}%
Comparing the coefficients of the Cartan-Weyl basis in (\ref{Lax}) one
deduces%
\begin{equation}
\dot{p}\cdot H=-\tsum_{\alpha \in \Delta }\alpha \cdot H~x_{\alpha
}x_{-\alpha }^{\prime }\qquad \text{and\qquad }\tsum_{\alpha \in \Delta
}(\alpha \cdot \dot{q})x_{\alpha }^{\prime }~E_{\alpha }=\tsum_{\alpha \in
\Delta }(p\cdot \alpha )x_{\alpha }^{\prime }~E_{\alpha }  \label{c2}
\end{equation}%
which are equivalent to the equations of motion $\dot{p}=-\nabla _{q}H$ and $%
\dot{q}=p$. In addition certain \textquotedblleft
unwanted\textquotedblright\ terms must cancel which leads to a functional
equation \cite{Cal5} \cite{OP0} for the as yet unspecified vector $z$. This
equation can be simplified using (\ref{xxV2}),%
\begin{eqnarray}
\alpha \cdot z &=&i\sum_{\substack{ \beta ,\gamma \in \Delta  \\ \alpha
=\beta +\gamma }}\varepsilon _{\beta ,\gamma }\frac{x_{\beta }x_{\gamma
}^{\prime }}{x_{\alpha }}=i\sum_{\beta \in \Delta }\varepsilon _{-\alpha
,\beta }~\frac{x_{\beta }x_{\alpha -\beta }^{\prime }-x_{\beta }^{\prime
}x_{\alpha -\beta }}{x_{\alpha }}  \label{c3} \\
&=&-2i\sum_{\beta \in \Delta }\varepsilon _{\alpha ,\beta }\frac{g_{\beta
}g_{\alpha +\beta }}{g_{\alpha }}V(\beta \cdot q)\ .  \notag
\end{eqnarray}%
Here we have used $g_{\beta }=g_{-\beta },\ V(u)=V(-u)$ as well as the
symmetries $\varepsilon _{\beta ,\gamma }=-\varepsilon _{\gamma ,\beta
}=\varepsilon _{-\alpha ,\beta }$ and $\varepsilon _{-\alpha ,\beta
}=-\varepsilon _{-\alpha ,\alpha -\beta }=-\varepsilon _{\alpha ,-\beta }$
for the structure constants. It is not clear \emph{a priori} that such a
vector $z$ always exists, but if it does, it must be unique as the
functional equation (\ref{c3}) applies among others also to the simple roots 
$\{\alpha _{i}\}$ which are linearly independent. Ambiguities arise when the
root space is realized as a hyperplane in a higher-dimensional space. Then
we might add an arbitrary vector $z^{\prime }$ which is orthogonal on $%
\Delta $, i.e. $z^{\prime }\cdot \alpha =0$ for all roots $\alpha $.
Employing the fundamental weights $\lambda _{i}$, which form the dual basis
to the simple roots, $\lambda _{i}\cdot \alpha _{j}=\delta _{ij}$, it is
immediate to derive that%
\begin{equation}
z(q)=-2i\sum_{\beta \in \Delta }V(\beta \cdot q)\sum_{i=1}^{\ell
}\varepsilon _{\alpha _{i},\beta }\frac{g_{\beta }g_{\alpha _{i}+\beta }}{%
g_{\alpha _{i}}}\lambda _{i}\ .  \label{z}
\end{equation}%
For an explicit computation of the vector $z$ and checking its consistency
with (\ref{c3}) for all roots $\alpha \in \Delta $ one needs to fix the
signs of the structure constants $\varepsilon _{\alpha ,\beta }$ in a
consistent manner. To this end we now specialize to a specific
representation and set $\ell =4$.

\subsection{The Lax pair in the vector representation of $A_{\ell }$}

Let $\{e_{i}\}_{i=1}^{5}$ be the orthonormal basis in the Euclidean space $%
\mathbb{R}^{5}$. Then a standard representation of the simple roots is \cite%
{Hum2}%
\begin{equation}
\alpha _{1}=e_{1}-e_{2},\quad \alpha _{4}=e_{2}-e_{3},\quad \alpha
_{3}=e_{3}-e_{4},\quad \alpha _{2}=e_{4}-e_{5}\ .  \label{rep}
\end{equation}%
Note that our labelling of the simple roots differs from the common one due
to our convention for the embedding (\ref{emb}). The entire root system
consists of the vectors%
\begin{equation}
\Delta =\Delta _{+}\cup -\Delta _{+},\qquad \Delta _{+}=\{\alpha
=e_{i}-e_{j}:1\leq i<j\leq 5\}
\end{equation}%
and the Cartan-Weyl basis is given in terms of the unit matrices $(\mathfrak{%
e}_{ij})_{lk}=\delta _{il}\delta _{jk}$ by identifying%
\begin{equation}
H_{i}=\mathfrak{e}_{ii}\text{\quad \quad and\quad \quad }E_{\alpha }=%
\mathfrak{e}_{ij}\text{\quad if\quad }\alpha =e_{i}-e_{j}\ .
\end{equation}%
For each simple root $\alpha _{i}$ one finds six non-vanishing structure
constants $\varepsilon _{\alpha _{i},\beta }$ only three of which are
independent due to the symmetries $\varepsilon _{\alpha _{i},\beta
}=-\varepsilon _{\beta ,\alpha _{i}}=\varepsilon _{-\alpha _{i}-\beta
,\alpha _{i}}=-\varepsilon _{\alpha _{i},-\alpha _{i}-\beta }$. Choosing $%
\beta $ to be positive the root decompositions $\alpha =\beta +\gamma $ and
the corresponding structure constants in (\ref{c3}) can be inferred from the
following table

\begin{center}
\begin{tabular}{|c|c|c|c|}
\hline\hline
& $(\beta _{1},\varepsilon _{\alpha _{i},\beta _{1}})$ & $(\beta
_{2},\varepsilon _{\alpha _{i},\beta _{2}})$ & $(\beta _{3},\varepsilon
_{\alpha _{i},\beta _{3}})$ \\ \hline\hline
$\alpha _{1}:$ & $(\alpha _{2}+\alpha _{3}+\alpha _{4},+1)$ & $(\alpha
_{3}+\alpha _{4},+1)$ & $(\alpha _{4},+1)$ \\ \hline\hline
$\alpha _{2}:$ & $(\alpha _{1}+\alpha _{3}+\alpha _{4},-1)$ & $(\alpha
_{3}+\alpha _{4},-1)$ & $(\alpha _{3},-1)$ \\ \hline\hline
$\alpha _{3}:$ & $(\alpha _{1}+\alpha _{4},-1)$ & $(\alpha _{4},-1)$ & $%
(\alpha _{2},+1)$ \\ \hline\hline
$\alpha _{4}:$ & $(\alpha _{2}+\alpha _{3},+1)$ & $(\alpha _{3},+1)$ & $%
(\alpha _{1},-1)$ \\ \hline\hline
\end{tabular}
\end{center}

In order to accommodate the possibility of two independent coupling
constants in the reduced Hamiltonian (\ref{newH02}) we need to identify the
subsets (\ref{DDD}). Setting $\Delta _{+}^{\prime }=\Delta ^{\prime }\cap
\Delta _{+}$ and $\Delta _{+}^{\prime \prime }=\Delta ^{\prime \prime }\cap
\Delta _{+}$ we have%
\begin{equation*}
\Delta _{+}^{\prime }=\{\alpha _{1},\alpha _{2},\alpha _{1}+\alpha
_{4},\alpha _{2}+\alpha _{3},\alpha _{3}+\alpha _{4}\}
\end{equation*}%
and%
\begin{equation*}
\Delta _{+}^{\prime \prime }=\{\alpha _{3},\alpha _{4},\alpha _{1}+\alpha
_{3}+\alpha _{4},\alpha _{2}+\alpha _{3}+\alpha _{4},\alpha _{1}+\alpha
_{2}+\alpha _{3}+\alpha _{4}\}\ .
\end{equation*}%
Making the same replacement in the dynamical variables (\ref{mu}) as in the
previous reduction of the Hamiltonian (\ref{newH0}) respectively (\ref%
{newH02}), we consider the crystallographic Lax pair with coordinates%
\begin{equation}
q\rightarrow \mu (q)=\tilde{\omega}(\tilde{q})=(\tilde{s}_{1},-\tilde{s}%
_{1}+\phi \tilde{s}_{2},\phi (\tilde{s}_{1}-\tilde{s}_{2}),-\phi \tilde{s}%
_{1}+\tilde{s}_{2},-\tilde{s}_{2})\ .  \label{redq}
\end{equation}%
Here $q=\sum_{i}q_{i}e_{i}$, i.e. the dynamical variables $q_{i}$ are the
coordinates with respect to the Euclidean basis $\{e_{i}\}$. The $\tilde{s}%
_{i}$, on the other hand, are the components with respect to the simple
roots, $\tilde{q}=\sum_{i}\tilde{s}_{i}\tilde{\alpha}_{i},$ and we have
inserted the explicit representation (\ref{rep}) for the simple roots $%
\alpha _{i}$. Note, however, that the dynamical variables $\tilde{q}_{i}$
are the ones with respect to the Euclidean basis. Both are related by a
simple linear transformation, i.e. $\tilde{s}_{i}$ is a linear function of
the Euclidean coordinates $\tilde{s}_{i}=\tilde{s}_{i}(\tilde{q}_{1},\tilde{q%
}_{2})$, whose explict form depends on the particular representation one
chooses for the non-crystallographic roots. For instance, if we set \cite%
{Hum2}%
\begin{equation}
\tilde{\alpha}_{1}=\tfrac{1}{\sqrt{2}}(\phi ,\sqrt{3-\phi })\qquad \text{and}%
\qquad \tilde{\alpha}_{2}=-\tfrac{1}{\sqrt{2}}(\phi ^{-1},\sqrt{3+\phi ^{-1}}%
)
\end{equation}%
then%
\begin{equation}
\tilde{s}_{1}=\tfrac{\phi }{\sqrt{2}}(\tilde{q}_{1}-\cot \tfrac{2\pi }{5}~%
\tilde{q}_{2})\qquad \text{and}\qquad \tilde{s}_{2}=\tfrac{1}{\sqrt{2}}(%
\tilde{q}_{1}-\cot \tfrac{\pi }{5}~\tilde{q}_{2})\ .
\end{equation}%
A similar replacement holds for the conjugate momenta, $p\rightarrow \mu (p)=%
\tilde{\omega}(\tilde{p})$. None of the algebraic properties of the Lax pair
is changed and one therefore verifies immediately that analogues of (\ref%
{L2H}) and (\ref{c2}) hold for the reduced, non-crystallographic
Hamiltonian. The latter imply the reduced crystallographic equations of
motion (\ref{eqm0}),%
\begin{equation}
\alpha \cdot \tilde{\omega}(\overset{\cdot }{\tilde{q}})~x^{\prime }(\omega
(\alpha )\cdot \tilde{q})=\alpha \cdot \tilde{\omega}(\tilde{p})~x^{\prime
}(\omega (\alpha )\cdot \tilde{q})\quad \Rightarrow \quad \tilde{\omega}(%
\overset{\cdot }{\tilde{q}})=\tilde{\omega}(\tilde{p})
\end{equation}%
and%
\begin{equation}
\tilde{\omega}(\overset{\cdot }{\tilde{p}})=\tsum_{\alpha \in \Delta }\alpha
~x_{\omega (\alpha )}x_{-\omega (\alpha )}^{\prime }=-\frac{g_{1}^{2}}{2}%
\tsum_{\alpha \in \Delta ^{\prime }}\alpha ~V^{\prime }(\alpha \cdot \tilde{%
\omega}(\tilde{q}))-\frac{g_{2}^{2}}{2}\tsum_{\alpha \in \Delta ^{\prime
\prime }}\alpha ~V^{\prime }(\alpha \cdot \tilde{\omega}(\tilde{q}))\ .
\end{equation}%
As discussed above a simple application of the map $\omega $ on both sides
then yields (\ref{eqm}). The non-trivial part of the reduction of the Lax
pair is the cancellation of the unwanted terms, i.e. solving the functional
equation (\ref{c3}). The root decomposition $\alpha =\beta +\gamma $ mixes
elements in the two subsets $\Delta ^{\prime }$, $\Delta ^{\prime \prime }$.
For instance we find for $\alpha =\alpha _{1}$,%
\begin{multline}
-\sum_{\beta \in \Delta }\varepsilon _{\alpha _{1},\beta }\frac{g_{\beta
}g_{\alpha _{1}+\beta }}{g_{\alpha _{1}}}V(\beta \cdot q)=  \notag \\
g_{2}\{V(q_{13})+V(q_{14})-V(q_{23})-V(q_{24})\}+g_{2}^{2}/g_{1}~%
\{V(q_{15})-V(q_{25})\}
\end{multline}%
with $q_{ij}=q_{i}-q_{j}$. Taking into account the reduced coordinates (\ref%
{redq}) one can solve (\ref{z}), but finds upon inserting the solution into (%
\ref{c3}) for arbitrary roots $\alpha $ that one is forced to set%
\begin{equation}
g_{1}=g_{2}=g\ .  \label{schade}
\end{equation}%
That is, one has $g_{\alpha }=g$ for all crystallographic roots $\alpha \in
\Delta $. With this restriction we recover after subtracting the superfluous
vector (since $z^{\prime }\cdot \alpha =0$ for all roots $\alpha $)%
\begin{equation}
z^{\prime }=\frac{e_{1}+\cdots +e_{5}}{5}~2ig\sum_{j=1}^{5}\sum_{k=1,k\neq
j}^{5}V(q_{j}-q_{k})
\end{equation}%
from (\ref{z}) the familiar solution of the $A_{\ell }$ series (see e.g. 
\cite{Pbook})%
\begin{equation}
z_{j}=2ig\sum_{k=1,k\neq j}^{5}V(q_{j}-q_{k}),
\end{equation}%
albeit in our case the coordinates $q_{i}$ have to be replaced by (\ref{redq}%
). Thus, we can conclude that the dynamical system defined through the
Hamiltonian (\ref{newH}) allows for a Lax pair formulation and hence the
quantities $I_{k}=\limfunc{Tr}L^{k}$ are conserved, i.e. $d(\limfunc{Tr}%
L^{k})/dt=0$. This is usually a strong indication that the model is indeed
Liouville integrable. It remains to show that the aforementioned integrals
of motion mutually Poisson commute and that at least $\tilde{\ell}=2$ of
them are non-vanishing and algebraically independent.

\subsection{Integrals of motion}

Having established the existence of the Lax operator $L$ we may now use its
explicit form,%
\begin{equation}
L=ig\left( 
\begin{array}{ccccc}
p_{1}/ig & x(q_{12}) & x(q_{13}) & x(q_{14}) & x(q_{15}) \\ 
x(q_{21}) & p_{2}/ig & x(q_{23}) & x(q_{24}) & x(q_{25}) \\ 
x(q_{31}) & x(q_{32}) & p_{3}/ig & x(q_{34}) & x(q_{35}) \\ 
x(q_{41}) & x(q_{42}) & x(q_{43}) & p_{4}/ig & x(q_{45}) \\ 
x(q_{51}) & x(q_{52}) & x(q_{53}) & x(q_{54}) & p_{5}/ig%
\end{array}%
\right) \ ,
\end{equation}%
to compute the integrals of motion. From examples of the crystallographic
Calogero-Moser models it is known that the algebraically independent
integrals of motion occur when the power $k$ of $I_{k}=\limfunc{Tr}L^{k}$
matches the degrees \{$d_{i}\}_{i=1}^{\ell }$ of the Coxeter group. For $%
A_{4}$ the degrees are $d_{i}=2,3,4,5$, while in our case of interest, $%
H_{2},$ they are $\tilde{d}_{i}=2,5$ \cite{Hum2}. For instance, in the case
of $A_{4}$ one verifies indeed that $I_{2},I_{3},I_{4},I_{5}$ are
algebraically independent, while for $I_{6}$ we have the relation%
\begin{equation}
I_{6}=-\frac{1}{8}~I_{2}^{3}+\frac{3}{4}~I_{2}I_{4}-\frac{1}{3}~I_{3}^{2}\ .
\end{equation}

One might anticipate that due to the additional dependencies in (\ref{redq})
some of the higher integrals of motion from the non-reduced $A_{4}$ theory
must become algebraically dependent in the reduction. For instance, consider
the integral of motion of degree three of the non-reduced model,%
\begin{equation}
I_{3}(p,q)=\limfunc{Tr}L^{3}(p,q)=\sum_{i}p_{i}^{3}+3g^{2}\sum_{i}p_{i}%
\sum_{j\neq i}V(q_{i}-q_{j})\ .
\end{equation}%
Since for the non-reduced as well as the reduced theory we have%
\begin{equation}
\limfunc{Tr}L=p_{1}+~...~+p_{5}=0,
\end{equation}%
one might expect the reduced integral of motion of degree three $I_{3}^{%
\text{red}}=I_{3}(\mu (p),\mu (q))$ to vanish. For purely algebraic reasons
we find the following simplification%
\begin{equation}
I_{3}^{\text{red}}=3g^{2}\sum_{i}\mu (p)_{i}\sum_{j\neq i}V(\mu (q)_{i}-\mu
(q)_{j})\ .
\end{equation}%
Because of the reduction (\ref{redq}) the purely kinetic term $\sum_{i}\mu
(p)_{i}^{3}$ is zero. The remainder, however, does not vanish in general.
The reason might be that the reduced set of crystallographic equations of
motion (\ref{eqm0}) is more restrictive than (\ref{eqm}) and that the
non-vanishing of this integral of motion is a remnant of the reduction
procedure. Consulting the literature we found that this issue is also not
addressed in the context of folding. It certainly requires a deeper
investigation of the model, for instance finding the explicit solutions of
the equations of motion. In comparison, the non-crystallographic reduction
of affine Toda field theories \cite{FKncATFT} involved two sets of
complementary degrees, whose union gives again the degrees of $A_{4}$. We
leave this question for future work.

In this context, it is also noteworthy, that the match between the degrees
of the Coxeter group and the powers of the integrals of motion, $k=d_{i}$,
appears to be an observation based on examples rather than a rigorous
mathematical theorem which applies to all known Calogero-Moser models. In
particular, as its verification depends on the explicit form of a given Lax
pair. For example, we followed for $H_{2}$ the Lax pair construction given
in \cite{Sas6}, which is based on the Coxeter group and the root system
alone; see the appendix. One verifies in the case of the root type
representation, as stated in \cite{Sas6}, that the Lax equation only holds
for the rational potential; see our earlier remarks in the introduction.
Explicit computation of the quantities $\limfunc{Tr}L^{k}$ shows that they
vanish for $k=1,3,5$. A similar analysis for the $A_{4}$ theory based on the
root type Lax pair gives an algebraically dependent expression for $\limfunc{%
Tr}L^{3}$. Thus, while one might expect the existence on an algebraically
independent integral of motion at a certain degree $d_{i}$, not every Lax
pair will provide one at the same power $k=d_{i}$.

\section{Conclusions}

The purpose of this article has been to extend a recent reduction procedure 
\cite{FKncATFT} based on the embedding of non-crystallographic Coxeter
groups into crystallographic ones to Calogero-Moser models. This lead us to
propose new integrable systems based on $H_{2,3,4}$ by extending from the
known rational potential to the trigonometric, hyperbolic and elliptic case.
The new feature has been the appearance of an additional term in the
potential energy, where the argument of the potential function is rescaled
by the golden ratio. It is this extra term which is difficult to identify in
the rational case, where it corresponds to a simple rescaling of the overall
coupling constant. As we have discussed in the text the additional potential
term can only be fully appreciated through the analysis of the underlying
structure of the Coxeter groups.

While there are mathematical differences between the embeddings (\ref{HADE})
and the folding \cite{DTfold} of a simply laced Lie algebra by a Dynkin
diagram automorphism, we motivated our procedure by pointing out
similarities and differences in the outcome. In particular, we showed that
the reduced crystallographic Hamiltonian, which originally depends only on
one coupling constant, still preserves invariance under the
non-crystallographic Coxeter group if a second, independent coupling
constant in front of the additional potential term is introduced. This is in
close analogy with folding, where one also starts from a Hamiltonian
incorporating only one coupling constant \cite{BST}, while the folded
Hamiltonian then allows for two independent couplings in front of the two
Weyl group orbits corresponding to short and long roots \cite%
{OP0,OP1,OP2,OP3}.

We addressed the question of Liouville integrability of the new models by
reducing the associated crystallographic Lax pair for the simplest case,
namely $H_{2},$ and found that (\ref{newH}) indeed possesses higher
conserved integrals of motion. We explained in detail where in the Lax pair
construction a consistency requirement forces one to set the two coupling
constants in front of the two potential terms equal. This does not preclude
the possibility that another Lax pair construction might be possible which
enables one to keep the two coupling constants independent, as it is the
case with the models based on non-simply laced Lie algebras. The only basis
on which to expect such an extension of the result found here, is the
Coxeter invariance of the Hamiltonian (\ref{newH2}). While the latter does
not necessarily imply Liouville integrability, it is the central, although
not exclusive, criterion for exact solvability. However, the models (\ref%
{newH2}) with $g_{2}=0$ have been previously discussed in the literature
from both conceptual points of view, Lax pairs \cite{Sas6} and exact
solvability \cite{HR4}, neither of these two approaches appears to have been
successful beyond the rational potential\footnote{%
In \cite{KS00} a complete proof of Liouville integrability is presented for
the classical Calogero-Moser models based on all root systems (including the
non-crystallographic ones) and for all type of potentials. This proof makes
use of the Lax pair formulation based on reflection operators of the Coxeter
group in \cite{Sas6}, which for the non-crystallographic groups only applies
to the rational potential; see e.g. the comment after equation (4.43) and
the solution (4.44) to the functional equation.}. This seems to indicate
that the second potential term in (\ref{newH}) is indeed essential.

Despite the fact that we have focussed for the Lax pair construction on the
simplest model only, it should be evident from this example calculation that
similar results are possible for the two other cases. In particular, as the
reduction procedure does not change any of the algebraic properties of the
crystallographic Lax pairs. Moreover, our analysis of the mathematical
structure associated with the embeddings (\ref{HADE}) presented in section 2
of this article, allows one to accommodate also other formulations of Lax
pairs \cite{Sas5,Sas6}; see the appendix. In this context it would be
interesting to verify whether also for these cases the two coupling
constants in (\ref{newH02}) have to be equal to ensure a consistent Lax pair.

To complete the proof of Liouville integrability one needs to verify that
sufficiently many of the conserved quantities $I_{k}$ are algebraically
independent and Poisson commute. The latter step can be carried out using
the concept of $r$-matrices \cite{STS} \cite{AT93,ABT94} \cite{Sk95} \cite%
{Braden:1997,Braden1998}. For the aforementioned reasons we expect that
similar constructions based on the crystallographic $r$-matrices will carry
through to the non-crystallographic models. What will be different is the
number of algebraically independent integrals of motion, only half as many
are needed due to the relation $\ell =2\tilde{\ell}$ for the ranks of the
two Coxeter groups in (\ref{HADE}). That such a decrease in the number of
independent integrals of motion occurs is to be expected from the change of
variables (\ref{mu}) in the reduction procedure, which introduces additional
dependencies. We postpone these more involved questions to future work.

Starting from the results in this article one can now proceed further and
discuss the corresponding quantum mechanical systems as well. In this
context new aspects arise, such as invariant theory and Dunkl operators \cite%
{Dunkl}. In the case of the rational quantum Calogero model it is customary
to add a confining harmonic potential in order to obtain a discrete
spectrum. Since we omitted this case from our discussion let us briefly
mention that the reduction procedure applies then as well and by a
computation along the same lines as the one in section 3, one finds that the
frequency of the harmonic oscillator is rescaled in the same manner as the
coupling constant of the crystallographic Hamiltonian, compare with (\ref%
{ggt}).

We conclude by emphasising once more the general nature of the reduction
from crystallographic to non-crystallographic Coxeter groups. Its possible
applications are as widespread as the one of the folding procedure \cite%
{DTfold}. The significant difference is that the non-crystallographic
reduction not only yields an alternative description but achieves to
maintain a connection with the theory of semi-simple Lie algebras whose rich
mathematical structure has found applications in many physical areas beyond
the one discussed here. A prominent example where this connection might be
relevant is the correspondence between Calogero-Moser models and
supersymmetric Yang-Mills theory in four dimensions, see e.g. \cite{DW96}%
.\bigskip 

\noindent \textbf{Acknowledgments}. We would like to thank Harry Braden and
Nenad Manojlovi\'{c} for discussions and providing various references. A.F.
would like to thank the Universidade do Algarve for kind hospitality. C.K.
is financially supported by a University Research Fellowship of the Royal
Society.

\begin{center}
\newpage \noindent \textbf{\large Appendix}
\end{center}

\appendix

\section{Comparison with the root type Lax pair based on Coxeter groups}

Our reduction of the Lie algebraic Lax pair for $A_{4}$ showed that the two
coupling constants in (\ref{newH02}) cannot be chosen independently.
However, there is the possibility that other construction schemes not based
on Lie algebras might be successful. If this would be the case we can set $%
g_{2}=0$ in (\ref{newH02}) and obtain a non-crystallographic Hamiltonian (%
\ref{newH}) where only the first potential term is present. These type of
models have been investigated in \cite{Sas6} by constructing Lax pairs based
on representations of the Coxeter group. Let us briefly describe this
alternative formulation in order to compare results. We shall concentrate on
the \textquotedblleft root-type\textquotedblright\ representation, see \cite%
{Sas5,Sas6} for other possibilities. The following description applies to
both the crystallographic and the non-crystallographic root systems.

Following \cite{Sas5,Sas6} we introduce a vector space $V_{\Delta }$ which
is the linear span of the following set of basis vectors $\{\left\vert
\alpha \right\rangle \}_{\alpha \in \Delta }$ labelled by the roots, i.e. $V$
is the direct sum $V_{\Delta }=\bigoplus_{\alpha \in \Delta }V_{\alpha }$
with $V_{\alpha }$ being the one-dimensional space spanned by $\left\vert
\alpha \right\rangle $. Obviously, the dimension of this space coincides
with the number of roots $\ell h$, with $h$ being the Coxeter number. The
Lax pair will be represented on this space. To formulate the latter, one
first introduces the following action of the Coxeter group,%
\begin{equation}
w\left\vert \alpha \right\rangle :=\left\vert w(\alpha )\right\rangle
,\qquad \text{for all }\alpha \in \Delta ,\;w\in \mathcal{W}\ .
\end{equation}%
As the root system is invariant under $\mathcal{W}$, so is $V_{\Delta }$.
With respect to this action the Weyl reflections $w=\sigma _{\alpha }$
assume the role of the step operators $E_{\alpha }$ in the previous Lie
algebraic formulation of the Lax pair. To mimic the Cartan elements $H_{i}$
one introduces the following set of linear operators \cite{Sas5,Sas6},%
\begin{equation}
h_{i}\left\vert \alpha \right\rangle :=\alpha ^{i}~\left\vert \alpha
\right\rangle ,\qquad \text{for all }\alpha \in \Delta ,\;i=1,...,\ell \ .
\end{equation}%
They commute among themselves, $[h_{i},h_{j}]=0$, and satisfy the crucial
relation \cite{Sas6}%
\begin{equation}
\lbrack h_{i},\sigma _{\alpha }]=\frac{2\alpha ^{i}}{\alpha ^{2}}~(\alpha
\cdot h)\sigma _{\alpha }
\end{equation}%
with the Weyl reflections. In addition one imposes a similar trace
convention as for the Cartan generators, $\limfunc{Tr}h_{i}h_{j}=~const~%
\delta _{ij}$. In terms of these operators and the Weyl reflections the Lax
pair now reads \cite{Sas6}%
\begin{equation}
L=p\cdot h+i\sum_{\alpha \in \Delta _{+}}g_{\alpha }x(\alpha \cdot
q)~(\alpha \cdot h)\sigma _{\alpha }\qquad \text{and}\quad \quad
M=i\sum_{\alpha \in \Delta _{+}}g_{\alpha }\frac{\alpha ^{2}}{2}~y(\alpha
\cdot q)\sigma _{\alpha }\ .
\end{equation}%
Similar to the calculation in the Lie algebraic construction one shows that
the Lax equation (\ref{Lax}) is equivalent to the equations of motion
provided $y=x^{\prime }$ and certain unwanted terms cancel. The latter
condition again leads to a functional equation which in the present
construction reads \cite{Sas6}%
\begin{equation}
\sum_{\alpha ,\beta \in \Delta _{R}}g_{\alpha }g_{\beta }\left[ \beta
^{2}x(\alpha \cdot q)y(\beta \cdot q)(\alpha \cdot \mu )-\alpha ^{2}y(\alpha
\cdot q)x(\beta \cdot q)(\sigma _{\alpha }(\beta )\cdot \mu )\right] =0\ .
\label{func2}
\end{equation}%
Here the set $\Delta _{R}$ contains all pairs $(\alpha ,\beta )$ of positive
roots for which $R=\sigma _{\alpha }\sigma _{\beta }\in \mathcal{W}$ is a
fixed rotation and $\mu =(\mu _{1},...,\mu _{\ell })$ can be any vector. The
advantage of this formulation is that it equally applies to crystallographic
and non-crystallographic Coxeter groups. However, for non-crystallograpic
systems the functional equation (\ref{func2}) is only satisfied for the
rational potential and the Lax construction breaks down for the
trigonometric, hyperbolic and elliptic case \cite{Sas6}.

\subsection{The case $H_{2}$}

This can be explicitly verified for the $H_{2}$ group by picking the
following representation of the roots \cite{Hum2},%
\begin{equation*}
\tilde{\beta}_{k}=\sqrt{2}(\cos \tfrac{\pi k}{5},\sin \tfrac{\pi k}{5}%
),\;k=1,2,...,10\ .
\end{equation*}%
In this representation the root set for a fixed rotation $R=\tilde{\sigma}_{%
\tilde{\beta}_{i}}\tilde{\sigma}_{\tilde{\beta}_{j}}$ is given by \cite{Sas6}%
\begin{equation*}
\tilde{\Delta}_{R}=\{(\tilde{\beta}_{i+k},\tilde{\beta}_{j+k}):k=0,1,...,4\}%
\ .
\end{equation*}%
One then verifies that the functional equation (\ref{func2}) does not hold
beyond the rational potential. For the conserved charges we find the
expressions%
\begin{equation*}
\limfunc{Tr}L=\limfunc{Tr}L^{3}=\limfunc{Tr}L^{5}=0\quad \text{and}\quad 
\frac{\limfunc{Tr}L^{2}}{10}=\frac{\limfunc{Tr}L^{4}}{15}=2H=\tilde{p}%
_{1}^{2}+\tilde{p}_{2}^{2}+\frac{25g^{2}(\tilde{q}_{1}^{2}+\tilde{q}%
_{2}^{2})^{4}}{(\tilde{q}_{1}^{5}-10\tilde{q}_{1}^{3}\tilde{q}_{2}^{2}+5%
\tilde{q}_{1}\tilde{q}_{2}^{4})^{2}}\ .
\end{equation*}%
Note the absence of a non-trivial charge for degree 5.

\subsection{The case $A_{4}$}

Here we choose the same representation of the roots as in the context of the
Lie algebraic Lax pair. The root type representation is now 20-dimensional.
Note that we have to shift the $h_{i}$-operators by a constant, $%
h_{i}\rightarrow h_{i}+1/\sqrt{10}$, in order to ensure the aforementioned
trace convention. This does not affect the commutation relations.

Now the root sets in the functional equation (\ref{func2}) involve at most
three pairs of roots. An example is 
\begin{equation*}
\Delta _{R}=\{(\alpha _{1},\alpha _{3}+\alpha _{4}),(\alpha _{1}+\alpha
_{3}+\alpha _{4},\alpha _{1}),(\alpha _{3}+\alpha _{4},\alpha _{1}+\alpha
_{3}+\alpha _{4})\},\qquad R=\sigma _{\alpha _{1}}\sigma _{\alpha
_{3}+\alpha _{4}}\ .
\end{equation*}%
Note the mixing of roots belonging to the subsets $\Delta ^{\prime },\Delta
^{\prime \prime }$ as in the previous construction of the Lax pair based on
the Lie algebra $A_{4}$. The functional equation (\ref{func2}) for the
trigonometric case with $x(u)=\cot u$ reads explicitly%
\begin{equation*}
g_{1}(g_{1}-g_{2})~\frac{\mu _{1}\sin 2q_{12}-\mu _{2}\sin 2q_{24}-2\mu
_{4}\cos q_{14}\sin (q_{12}-q_{24})}{2\sin ^{2}q_{12}\sin ^{2}q_{24}}=0\ .
\end{equation*}%
From this and similar equations we again infer that we need to set $%
g_{1}=g_{2}$ to satisfy the Lax equation.

\end{document}